%% file: manuscript_TSP-acm.tex
\documentclass[sigconf]{acmart}

\AtBeginDocument{%
  }

\copyrightyear{2026}
\acmYear{2026}
\setcopyright{cc}
\setcctype{by}
\acmConference[PEARC '26]{Practice and Experience in Advanced Research Computing}{July 26--30, 2026}{Minneapolis, MN, USA}
\acmBooktitle{Practice and Experience in Advanced Research Computing (PEARC '26), July 26--30, 2026, Minneapolis, MN, USA}
\acmDOI{10.1145/3785462.3815790}
\acmISBN{979-8-4007-2377-3/2026/07}

\usepackage{amsmath}
\usepackage{subcaption} 
\usepackage{cuted} 
\usepackage{hyperref}

\begin{document}

\title{A Quantitative Framework for Comparing Classical and Quantum Algorithms for the Traveling Salesman Problem}

\author{Krit Grover}
\email{krit.grover@mail.utoronto.ca}
\author{Marcelo Ponce}
\email{m.ponce@utoronto.ca}
\affiliation{%
  \institution{Department of Computer and Mathematical Sciences, University of Toronto Scarborough}
  \city{Toronto}
  \state{ON}
  \country{Canada}
}


\input{abstract}




\maketitle


\input{./TSP-report}

\input{./ack}

\bibliographystyle{ACM-Reference-Format}
\bibliography{refs}

\end{document}

%% file: abstract.tex
\begin{abstract}
The Traveling Salesman Problem is a classical NP-hard problem
with significant implications in logistics, circuit design, and
operations research. This paper presents a comparative study of
four approaches to solving the Traveling Salesman Problem:
brute-force enumeration, a 2-approximation algorithm using minimum spanning trees,
simulated annealing, and the Quantum Approximate Optimization Algorithm.
We implement each technique and evaluate them on graphs of varying
sizes to analyze performance, solution quality, and scalability.
In doing so, we have also developed an open-source framework that
allows researchers and practitioners to explore, test and extend these methods.
\end{abstract}

%% file: TSP-report.tex
\section{Introduction}
\label{sec:intro}

The Traveling Salesman Problem (TSP) is one of the most well-known and extensively studied challenges in combinatorial optimization. The problem statement is the following: given a list of cities and a starting city, what is the shortest possible route for a salesman to visit all cities exactly once and return to the starting point? Despite its simple formulation, the TSP is NP-hard, and exact solutions are computationally infeasible for large inputs due to the factorial growth of possible routes. Over the decades, the problem has evolved into a benchmark for evaluating optimization methods across computer science, operations research, and now, quantum computing.

Classical techniques such as simulated annealing, and genetic algorithms have demonstrated scalability and near-optimal performance, especially on large instances \cite{qaoa}. However, they do not guarantee optimality and often require careful parameter tuning. Simultaneously, advancements in quantum computing have motivated exploration into algorithms like the Quantum Approximate Optimization Algorithm (QAOA) \cite{qaoa} and novel state encoding schemes that promise polynomial speed-ups for certain problems. However, most current quantum methods are constrained by noise, and limited qubit counts, even for TSP instances with fewer than ten cities.

These studies often emphasize either algorithmic theory or abstract numerical benchmarks, with limited focus on implementation-level trade-offs, particularly when comparing quantum and classical solvers in a unified experimental framework. Although multiple papers \cite{comparativestudyquantum} compare approximation ratios or simulation fidelity, fewer works have taken a hands-on, reproducible, and empirical approach to study the performance, scalability, and resource requirements of classical and quantum TSP solvers under consistent constraints.

The purpose of this research is to fill that gap by conducting a comparative implementation study and provide an open source framework for four distinct TSP-solving approaches:
\begin{itemize}
    \item Brute-force Search
    \item Minimum Spanning Tree $2$-Approximation (MST)
    \item Simulated Annealing (SA)
    \item Quantum Approximate Optimization Algorithm (QAOA)
\end{itemize}

We evaluated each method based on performance metrics such as execution time, solution quality, and scalability with varying problem sizes. In particular, we focus on the trade-offs encountered during implementation, from memory usage to algorithmic bottlenecks, and highlight the engineering decisions made to optimize or adapt each approach for practical use.

Although the TSP is often used as a flagship example to illustrate the potential of quantum computing, recent critical reviews have challenged its suitability as a benchmark for demonstrating near-term quantum advantage \cite{smithmiles}. A central argument is that popular quantum methods, such as QAOA and Quantum Annealing, require reformulating the TSP as a Quadratic Unconstrained Binary Optimization problem. This formulation creates a complex and difficult optimization landscape and is not competitive with the methods used by the latest classical solvers \cite{smithmiles}. Our research addresses this debate by conducting a hands-on, empirical comparison of classical and quantum solvers, focusing on the very implementation-level trade-offs and scalability challenges that are central to this discussion.

This paper is organized as follows. Sec.~\ref{sec:methods} provides a review of existing approaches to solving the TSP, and describes our methodology and implementation. Sec.~\ref{sec:results} presents our experimental results and comparative analysis across the different methods. Finally, Sec.~\ref{sec:concl} concludes the paper with key takeaways, limitations of our study, and potential directions for future work.

\section{Methods}
\label{sec:methods}

\subsection*{Brute Force Approach}

The brute force approach represents the most straightforward method for solving the TSP, examining all possible permutations of city visits to identify the optimal route. While this guarantees finding the global optimum, for an $n$-city problem, the brute force method requires evaluating $n!$ potential solutions, which quickly becomes prohibitive and limits its practical application to small problem instances (typically $n \leq 10$ cities). This implementation serves as a crucial baseline for validating the correctness of heuristic and quantum approaches, as it guarantees optimal solutions.

\subsection*{MST $2$-Approximation}

The MST $2$-approximation algorithm offers a polynomial-time approach to the TSP, providing solutions guaranteed to be within a factor of $2$ of the optimal solution. This method constructs a minimum spanning tree of the graph and then performs a preorder traversal to create a Hamiltonian cycle. The MST approach has been extensively studied since its introduction by Held and Karp in $1970$ \cite{mst}, who demonstrated its effectiveness for obtaining near-optimal solutions efficiently. While it does not guarantee optimality, the MST $2$-approximation provides a reasonable balance between solution quality and computational efficiency, particularly for larger problem instances \cite{qaoa}.

\subsubsection*{Implementation and Design Decisions for MST-based TSP}
The implementation follows the canonical MST-based 2-approximation algorithm:
\begin{enumerate}
    \item \textbf{MST Construction:} Use Prim's algorithm to find the minimum spanning tree of the input graph.
    \item \textbf{Tree Representation:} Convert the MST edge list into an adjacency list for efficient traversal.
    \item \textbf{DFS Traversal:} Perform depth-first search preorder traversal starting from node 0.
    \item \textbf{Cycle Formation:} Create the Hamiltonian cycle using the DFS ordering and return to the starting node.
\end{enumerate}
This structured approach separates concerns clearly, making each algorithmic step independently verifiable and modifiable.

\subsubsection*{Trade-offs and Scalability of the MST-Based TSP Approach}
A central trade-off is the requirement that the input graph be fully connected. The algorithm generates a TSP tour by performing a pre-order traversal of the MST and connecting each consecutive node in the traversal order to form a cycle. If the graph is not fully connected, it is possible that, after traversing one subtree (for example, the left subtree), there is no edge linking the last node of that subtree to the next node in another subtree (such as the right subtree). In such cases, the resulting path will be invalid or incomplete. This limitation means the MST approach is best suited to complete graphs, or to graphs where missing edges are artificially assigned large weights to ensure connectivity, but that distorts the practical meaning of the solution.

From a scalability perspective, the MST-based approach is highly efficient. The construction of the MST using the Prim's or Kruskal's algorithm runs in $\mathcal{O}(E.\log V)$ time, with $E$ being the number of edges and $V$ the number of nodes, and the subsequent DFS traversal and cycle formation are linear in the number of nodes. This allows the method to scale to large problem instances that are intractable for exact algorithms such as brute-force search.

\subsection*{Simulated Annealing}

Simulated annealing emerged as a powerful metaheuristic for addressing the TSP following the groundbreaking work of Kirkpatrick, Gelatt, and Vecchi in $1983$ \cite{simulatedannealing}. Drawing inspiration from the physical annealing process in metallurgy, SA avoids becoming trapped in local minima through a probabilistic acceptance criterion that allows occasional uphill moves, particularly in the early stages of the optimization process \cite{simulatedannealing,practicalconsiderationsa}. The so-called ``temperature" is a control parameter that governs the probability of accepting higher-cost tours, typically decreasing according to a predefined cooling schedule to gradually shift the algorithm from global exploration to local refinement.

At ``high temperatures", the algorithm is more likely to accept worse solutions, allowing it to escape local minima. As the temperature decreases, the acceptance probability for worse solutions decreases, focusing the search on increasingly promising regions. This balance between exploration and exploitation makes SA particularly effective for complex problems with many local optima.

\subsubsection*{Implementation}

Our SA implementation extends the typical canonical structure:
\begin{enumerate}
    \item \textbf{Initialization:} Instead of relying solely on random shuffles (which often fail on sparse graphs), a multi-strategy approach was used, that consisted of using greedy nearest-neighbor heuristics, random valid connections, backtracking and restarts and using pure randomization as a last resort.
    \item \textbf{Neighbor Generation:} Targeted swaps based on graph topology using adjacency lists and $n$-hop neighbors were used instead of simple random swaps, improving both solution quality and convergence speed. Specifically, when generating a new candidate solution, the algorithm prioritizes swapping cities that are topologically connected (adjacent or within $n$ hops in the graph structure) rather than randomly selecting any pair of cities, leading to more meaningful perturbations.
    \item \textbf{Acceptance Criterion:} Always accept better solutions; accept worse solutions with a probability based on the temperature and cooling rate.
    \item \textbf{Cooling Schedule:} Use an exponential decay to gradually reduce the temperature, controlling the transition from exploration to exploitation.
    \item \textbf{Early Stopping:} Terminate if the temperature drops below a threshold or if no improvement is seen for a fixed number of iterations.
\end{enumerate}

The SA algorithm for TSP involves two key operations: perturbing the current solution and evaluating its quality. As described in \cite{simulatedannealing}, ``the main advantage of SA is its simplicity. The cooling schedule plays a crucial role in SA performance, with geometric cooling and polynomial-time cooling being the most common approaches \cite{simulatedannealing,practicalconsiderationsa}. Practical implementations of SA for the TSP have demonstrated impressive results, with solutions for instances involving thousands of cities achievable within reasonable computation times \cite{simulatedannealing}.

\subsubsection*{Implementation Trade-offs and Scalability}
SA can produce high-quality near-optimal tours with a well designed cooling schedule. A schedule that cools too rapidly may lead to premature convergence on suboptimal solutions, while too slow cooling increases computation time exponentially. For TSP instances with hundreds or thousands of cities, SA's iterative neighborhood search can scale reasonably, with time complexity often $\mathcal{O}(N^2)$ per iteration where $N$ is the number of cities, but the total runtime depends on the number of iterations, which grows with problem size to maintain quality \cite{simulatedannealing}. Additionally, initialization strategies, such as greedy heuristics for starting tours, add upfront complexity but improve convergence; yet, poor initialization can increase scalability problems by requiring more annealing cycles.

\subsection*{Quantum Approximate Optimization Algorithm}
The Quantum Approximate Optimization Algorithm (QAOA) is a variational hybrid quantum-classical algorithm designed to solve combinatorial optimization problems, like the TSP. QAOA is deeply connected with adiabatic quantum computation, but approximates the adiabatic process by parameterizing the infinitely long time evolution into finite time steps. 

\subsubsection*{Implementation}
Our implementation of QAOA for the TSP is fundamentally guided by the methodology proposed in \cite{qaoa}, which introduces a resource-efficient edge-to-qubit problem mapping and a novel approach to handling constraints, as opposed to the approach used by IBM Q that requires twice the number of qubits and the additional complexity of penalty terms \cite{qaoa, ibmq2018}. In this approach, a computational basis state $|x\rangle$ represents a subgraph where the $k$-th qubit is $|1\rangle$ if the edge is included in the tour and $|0\rangle$ otherwise. Table~\ref{tab:qaoa-comparison} summarizes the key differences between the two 
approaches.

\begin{table*}[t]
\begin{tabular}{lcc}
\hline
\textbf{Property} & \textbf{IBM~Q \cite{ibmq2018} (Node-based)} & \textbf{Ruan et al. \cite{qaoa} (Edge-based)} \\
\hline
Qubit count          & $n^2$              & $n(n{-}1)/2$           \\
Encoding scheme      & City--position pairs & Edges                 \\
Constraint handling  & Penalty terms in $H_C$ & Encoded in $H_B$   \\
Feasibility guarantee & No (penalty-dependent) & Yes (by construction) \\
Mixer Hamiltonian    & Standard $X$-mixer & Feasibility-preserving  \\
Circuit depth (per layer) & $O(n^4)$ penalty terms & $O(|\mathcal{S}|^2)$ transitions \\
Initial state        & $|+\rangle^{\otimes n^2}$ & Superposition over valid cycles \\
\hline
\end{tabular}
\caption{Comparison of the IBM~Q \cite{ibmq2018} node-based encoding and the Ruan et al.\cite{qaoa} edge-based encoding for QAOA applied to TSP.}
\label{tab:qaoa-comparison}
\end{table*}

\textbf{Hamiltonian Formulation : }The cost Hamiltonian $H_C$ encodes the cost function, which in our case, measures the tour cost. By mapping binary variables $x_{uv} \in \{0,1\}$ to Pauli-$Z$ operators via $x_{uv} \to (I - Z_{uv})/2$, $H_C$ takes the form:\begin{equation}H_C = \sum_{(u,v) \in E} w_{uv} \left( \frac{I - Z_{uv}}{2} \right),\end{equation}where $w_{uv}$ is the weight of the edge between cities $u$ and $v$.

The mixing Hamiltonian $H_B$ drives transitions between different candidate solutions to explore the search space. In our implementation, $H_B$ is constructed to drive transitions only between feasible solutions $|x\rangle$ and $|x'\rangle$ that are separated by a 2-opt move (a Hamming distance of 4 in the edge basis). Mathematically, this is defined as a graph state walk:\begin{equation}H_B = \sum_{\substack{x, x' \in \mathcal{S} \ d(x,x')=4}} \left( |x\rangle\langle x'| + |x'\rangle\langle x| \right).\end{equation}This ensures that the quantum state evolution remains strictly within the subspace of valid Hamiltonian cycles, eliminating the need for penalty terms.

The algorithm operates by preparing the initial state $|s\rangle$ as an equal superposition of all valid Hamiltonian cycles,
\begin{equation}
|s\rangle = \frac{1}{\sqrt{|\mathcal{S}|}} \sum_{x \in \mathcal{S}} |x\rangle.
\end{equation}and then applying a sequence of unitary operations.  The parameterized ansatz state at depth $p$ is defined as:
\begin{equation}
|\psi(\vec{\gamma}, \vec{\beta})\rangle = \prod_{k=1}^p e^{-i\beta_k H_B} e^{-i\gamma_k H_C} |s\rangle,
\end{equation} where $H_C$ is the problem Hamiltonian, and $H_B$ is the mixing Hamiltonian. 
A classical optimizer iteratively updates the parameters $(\vec{\gamma}, \vec{\beta})$ to minimize the expectation value $E(\vec{\gamma}, \vec{\beta}) = \langle \psi | H_C | \psi \rangle$.

Comparative studies have shown that different designs of QAOA mixers, such as the X mixer, the XY mixer, and the row switch (RS) mixer, exhibit varying performance characteristics in terms of the approximation ratio, the percentage of true solutions, and the computational resource requirements \cite{comparativestudyquantum}. These findings highlight the importance of selecting appropriate quantum approaches based on the characteristics of specific problems and the available resources \cite{comparativestudyquantum}.

Our implementation follows the Ruan et al. \cite{qaoa} formulation but differs in several practical 
respects. First, we implement the full pipeline in Python using PennyLane \cite{bergholm2022pennylaneautomaticdifferentiationhybrid} as the quantum 
simulation backend, whereas the original work used Mathematica for numerical simulation \cite{qaoa}. 
Second, we embed the QAOA solver within a unified benchmarking framework alongside 
three classical baselines, 
enabling direct comparisons of execution time, solution quality, and memory 
usage under consistent experimental conditions. Third, our study provides  
measurements of the classical precomputation bottleneck, specifically the cost of 
enumerating all Hamiltonian cycles to construct $|s\rangle$ and $H_B$, which was not 
quantified in the original theoretical treatment \cite{qaoa}. These measurements directly inform 
the scalability analysis presented in Sec.~\ref{sec:results}.

\textbf{Time Evolution: Exact vs.\ Trotterized}:
Each QAOA layer applies the unitaries $e^{-i\gamma_k H_C}$ and $e^{-i\beta_k H_B}$.
In our PennyLane implementation, these operators can be realized in two ways.
Setting \texttt{num\_approx}\,$=0$ applies \texttt{qml.evolve}, which evaluates the matrix exponential of each Hamiltonian during state-vector simulation.
For \texttt{num\_approx}\,$\geq 1$, we instead use \texttt{qml.ApproxTimeEvolution}, which implements a first-order Trotter--Suzuki decomposition of the evolution into $n$ steps:
\begin{equation}
e^{-i H t} \approx \prod_{k=1}^{n} \prod_{j} e^{-i H_j t/n},
\end{equation}
where $H_j$ are the Pauli terms of the Hamiltonian and $n$ is the \texttt{num\_approx} parameter passed to the circuit.
This approximation becomes exact when all terms commute, but for general Hamiltonians the error decreases as $n$ increases at the cost of a deeper circuit and longer simulation time per QAOA layer.
Both $H_C$ and $H_B$ contain non-commuting terms: $H_C$ from weighted Pauli-$Z$ operators on distinct edges, and $H_B$ from four-qubit Pauli strings associated with 2-opt transitions, so the choice of $n$ is a practical engineering knob rather than a purely theoretical detail.
The exact-evolution path is feasible only in classical simulation; on hardware, Trotterization (or an equivalent gate decomposition) would be required regardless.

\subsubsection*{Trade-offs and Scalability}
The implementation of QAOA within a classical simulation environment introduces significant trade-offs and scalability challenges.

\begin{enumerate}
    \item \textbf{Classical Pre-computation Bottleneck:} The most severe limitation is the classical precomputation step required to identify all Hamiltonian cycles to initialize $|s\rangle$ and construct $H_B$. This exhaustive search has a factorial time complexity relative to the number of cities, making the entire approach intractable for graphs larger than a few nodes.
    \item \textbf{Simulation Memory Cost:} Simulating the quantum system is itself exponentially expensive. The state vector required for the simulation has a size of $2^N$, where $N$ is the number of qubits that is $n(n-1)/2$ for a $n$ city graph. This memory requirement grows rapidly (see Fig.~\ref{fig:memory_comparison}), making simulations for even moderately sized TSPs infeasible on classical hardware.
    \item \textbf{Approximation Quality vs. Circuit Depth:} As with any QAOA implementation, the quality of the solution depends on the number of layers, $p$. A larger $p$ allows the algorithm to explore the solution space more thoroughly and can lead to better approximation ratios. However, this comes at the cost of a deeper quantum circuit, more parameters for the classical optimizer to handle, and longer simulation times.
    \item \textbf{Trotterization Fidelity vs. Gate Depth:} When using QAOA, the per-layer Trotter step count \texttt{num\_approx} provides an additional depth knob. Low values ($n=1$) minimize gate count but can introduce sufficient unitary error to degrade optimized tour quality; higher values improve fidelity but multiply the number of Pauli rotations required for each Hamiltonian, compounding the cost of the already dense $H_B$ mixer.
    \item \textbf{Exact vs.\ Trotterized Scalability Ceiling:} The two evolution modes hit different practical limits in classical simulation. Exact evolution is not merely a question of available RAM. \texttt{qml.evolve} must form the full time-evolution operator from the Hamiltonian, via matrix exponentiation and subsequent decomposition into circuit operations, and this step becomes computationally prohibitive as the number of qubits and Hamiltonian terms grow. In our experiments on Trillium nodes with ${\sim}800$~GB of memory per dual-socket allocation, exact evolution (\texttt{num\_approx}\,$=0$) failed for instances beyond five cities due to overflow in the underlying decomposition, not because the state vector exceeded available memory. Trotterized evolution avoids materializing the full unitary at once and, in our benchmarks, remained runnable up to seven cities.
\end{enumerate}

\subsection{An Open-Source Framework}
\label{sec:repo}

All the methods described before were implemented in Python and can be
freely accessed, as well as details about their technical implementation
and how to use them, in a public open-source repository:
\url{https://github.com/kritgrover/tsp-framework}. One of the goals of this work, is to make this ready-to-use framework
available for researchers, practicioners,  and educators.
In this way we aim to facilitate and speed-up the discovery and education
process, as well as adhering to the principles of reproducible research.

\section{Results}
\label{sec:results}

This section presents the empirical results of our comparative study. We evaluated the four implemented TSP solvers on graphs of varying sizes. The evaluation focuses on three aspects: execution time, solution quality and memory usage. For these benchmarks, all the graphs are fully connected and we set some default parameters for the QAOA implementation and Simulated Annealing. For QAOA, we set layers $= 2$ and optimization steps $= 100$. For Simulated Annealing, we set initial temperature $= 2000$ and cooling rate $= 0.03$.
For the QAOA time-evolution study, we additionally swept the Trotter step count \texttt{num\_approx} over $\{0, 1, 2, 3, 4\}$, where $0$ selects exact \texttt{qml.evolve} and values $1$--$4$ select \texttt{qml.ApproxTimeEvolution} with the corresponding number of steps per Hamiltonian per layer. Exact-evolution runs were limited to at most five cities; Trotterized runs extended to seven cities, reflecting the different scalability ceilings of the two backends rather than a memory constraint on the host machine.
Numerical simulations were run in the Trillium supercomputer\footnote{\url{https://docs.alliancecan.ca/wiki/Trillium_Quickstart}}
hosted at SciNet HPC Consortium at the University of Toronto.
The compute node consists of 192 cores from two 96-core AMD EPYC 9655 CPUs ("Zen 5" a.k.a. "Turin") at 2.6 GHz (base frequency), each CPU has 755~GiB/810~GB of available memory.

\subsection{Execution Time vs. Problem Size}

\begin{figure}[h]
    \centering
    \includegraphics[width=\linewidth]{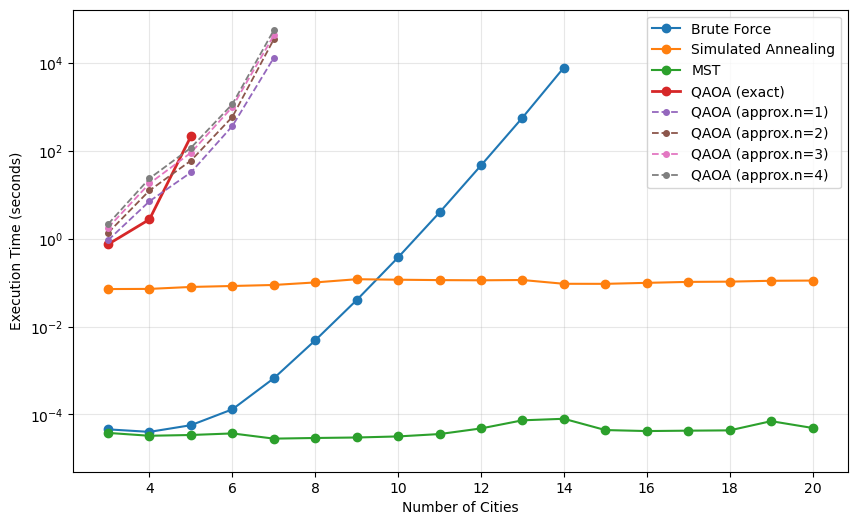}
    \caption{Execution time (log scale) versus number of cities (fully connected) for each algorithm.}
    \label{fig:exec_time}
\end{figure}

To assess scalability, we measured the execution time of each algorithm as a function of the number of cities. The results are presented in Fig.~\ref{fig:exec_time}.

The plot clearly illustrates the fundamental divide in computational complexity between the different approaches.

\begin{enumerate}
    \item \textbf{Brute-Force:} The execution time of the brute force algorithm exhibits exponential growth, confirming its factorial time complexity.
    \item \textbf{MST Approximation and Simulated Annealing:} In stark contrast, the MST and Simulated Annealing algorithms demonstrate exceptional performance and scalability. Their execution times remain low and grow modestly across the entire range of problem sizes tested (up to $n=20$), highlighting the efficiency of polynomial-time and heuristic approaches for larger problem instances.
    \item \textbf{QAOA:} Like brute force, QAOA relies on a classical precomputation of all valid Hamiltonian cycles and shows a sharp increase in runtime as problem size grows, limiting its practical reach in simulation. With relatively few data points, the timing results still reveal the expected behaviour across evolution backends: for Trotterized runs, the execution time increases with increasing \texttt{num\_approx}, whereas exact evolution escalates much more steeply and could not be extended beyond five cities. In either case, QAOA remains orders of magnitude slower than the classical methods.
\end{enumerate}

\subsection{Solution Quality vs Problem Size}

\begin{figure}[h]
    \centering
    \includegraphics[width=\linewidth]{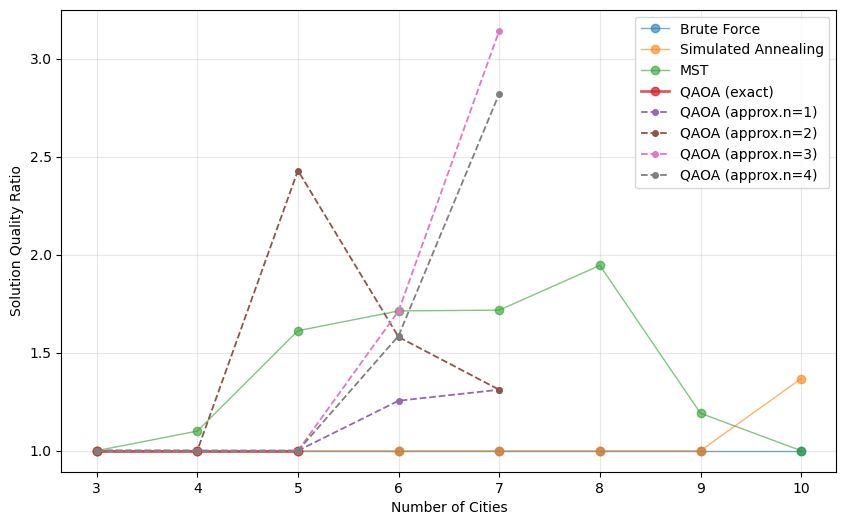}
    \caption{Approximation ratios ($C_{\text{alg}}/C_{\text{opt}}$) versus number of cities (fully connected) for each algorithm on instances where the optimum was available from brute force.}
    \label{fig:approx_ratios}
\end{figure}

Beyond performance, we evaluated the quality of the solutions produced by the approximation and heuristic algorithms. We computed the \textit{approximation ratio} (calculated as the tour cost / optimal cost) for each method, where a value closer to $1.0$ indicates a solution closer to optimal, and higher values indicate poorer solution quality. The optimal cost was determined by our brute-force solver. The results are shown in Fig.~\ref{fig:approx_ratios}.

\begin{enumerate}
    \item \textbf{Simulated Annealing:} Its approximation ratio remained very close to 1.0 across all tested problem sizes, highlighting the effectiveness of SA in navigating complex solution spaces to find high-quality solutions without requiring exhaustive search.

    \item \textbf{MST 2-Approximation:} Although the approximation ratio remained well within its theoretical upper bound of 2.0, its practical performance fluctuated significantly. This demonstrates the trade-off inherent to the MST approach: it provides a fast solution with a worst-case quality guarantee, but its typical performance is not as consistently near-optimal as SA.

    \item \textbf{QAOA:} On the small instances where QAOA could be completed, exact evolution yielded near-optimal tours with 4 and 5 cities, with only minor variation across graph instances and topologies. Because classical precomputation limited exact-evolution runs to at most five cities, we cannot yet characterize scaling behavior conclusively; the available data are nonetheless consistent with the broader challenges of applying near-term variational quantum algorithms to combinatorial problems. By contrast, Trotterized evolution showed larger scatter in approximation ratio, as expected from an approximate Suzuki decomposition whose fidelity depends on the Trotter step count. Neither increasing \texttt{num\_approx} nor increasing the number of cities produced a clear monotonic trend in solution quality within our benchmark window, suggesting that Trotter discretization error and fixed circuit depth ($p=2$) overshadow any emergent scaling signal. More systematic sweeps will be needed to determine whether quality improves predictably as decomposition fidelity increases.
\end{enumerate}

\subsection{Memory Usage vs Problem Size}

\begin{figure}[h]
    \centering
    \includegraphics[width=\linewidth]{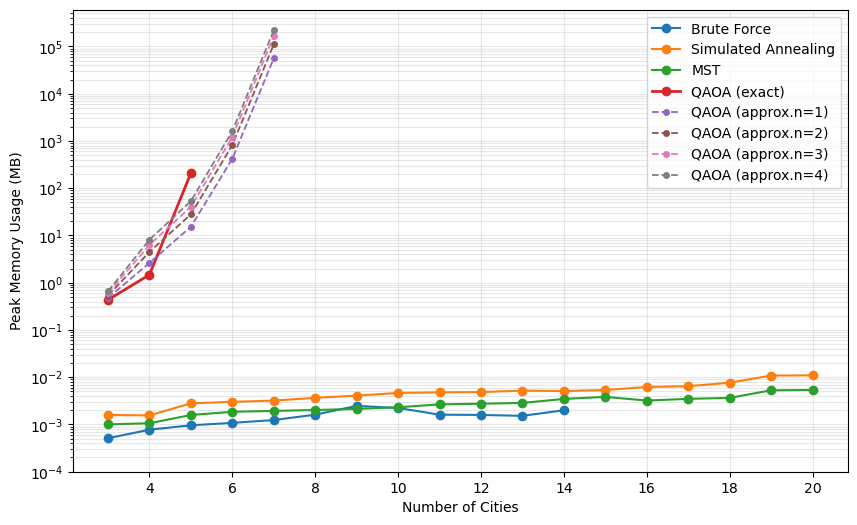}
    \caption{Peak memory usage (log scale) versus number of cities (fully connected) for each algorithm.}
    \label{fig:memory_comparison}
\end{figure}

Finally, we measured peak memory consumption as a function of $n$ to capture the space overhead of each approach.
Fig.~\ref{fig:memory_comparison} reports peak memory usage on a logarithmic $y$-axis.

Fig.~\ref{fig:memory_comparison} shows that memory consumption separates the QAOA pipeline from the purely classical baselines by several orders of magnitude:
\begin{enumerate}
    \item \textbf{Classical methods:} Across the tested range, the classical algorithms operate in a low-memory regime in this implementation. MST and SA show a gradual increase in peak memory as $n$ grows, consistent with storing distance/cost information and maintaining candidate tours or auxiliary structures whose size grows polynomially with $n$. Brute force exhibits similarly small peak memory within the limited range where it was executed ($n \leq 15$), with the primary bottleneck for brute force manifesting in time rather than memory at these sizes.
    
    \item \textbf{QAOA:} In stark contrast, QAOA exhibits a dramatic increase in peak memory even over very small problem sizes, rising from approximately $1$MB at 3 cities to $10^{2}$MB at 5 cities. This sharp jump reflects the exponential state-vector footprint ($2^N$ amplitudes for $N = n(n{-}1)/2$ qubits) together with substantial classical overhead from Hamiltonian-cycle enumeration and the data structures maintained during variational optimization. Comparing the two evolution backends, exact evolution grows faster with increasing cities, because \texttt{qml.evolve} must materialize the full time-evolution operator via matrix exponentiation and decomposition before simulation proceeds. Trotterized evolution follows a gentler upward trend, since \texttt{qml.ApproxTimeEvolution} applies the Hamiltonian term-by-term without constructing the complete unitary at once. Within the Trotterized runs, peak memory also increases with increasing \texttt{num\_approx}, as expected: a larger step count deepens each QAOA layer and expands the intermediate state and gate bookkeeping required during simulation. More broadly, these results highlight a key practical constraint for near-term quantum workflows evaluated via classical simulation: the memory footprint can become prohibitive well before reaching problem sizes of real interest.

\end{enumerate}

\subsection{A 5-City Case Study}
To provide a more concrete illustration of the solution quality trade-offs discussed previously, we present a visual comparison of the final tours generated by each of the four algorithms on a specific 5-city graph instance. This case study highlights the practical differences in the paths identified by exact, heuristic, and quantum-inspired methods.

\begin{figure*}
    \centering
    \begin{subfigure}{0.48\textwidth}
        \centering
        \includegraphics[width=\linewidth]{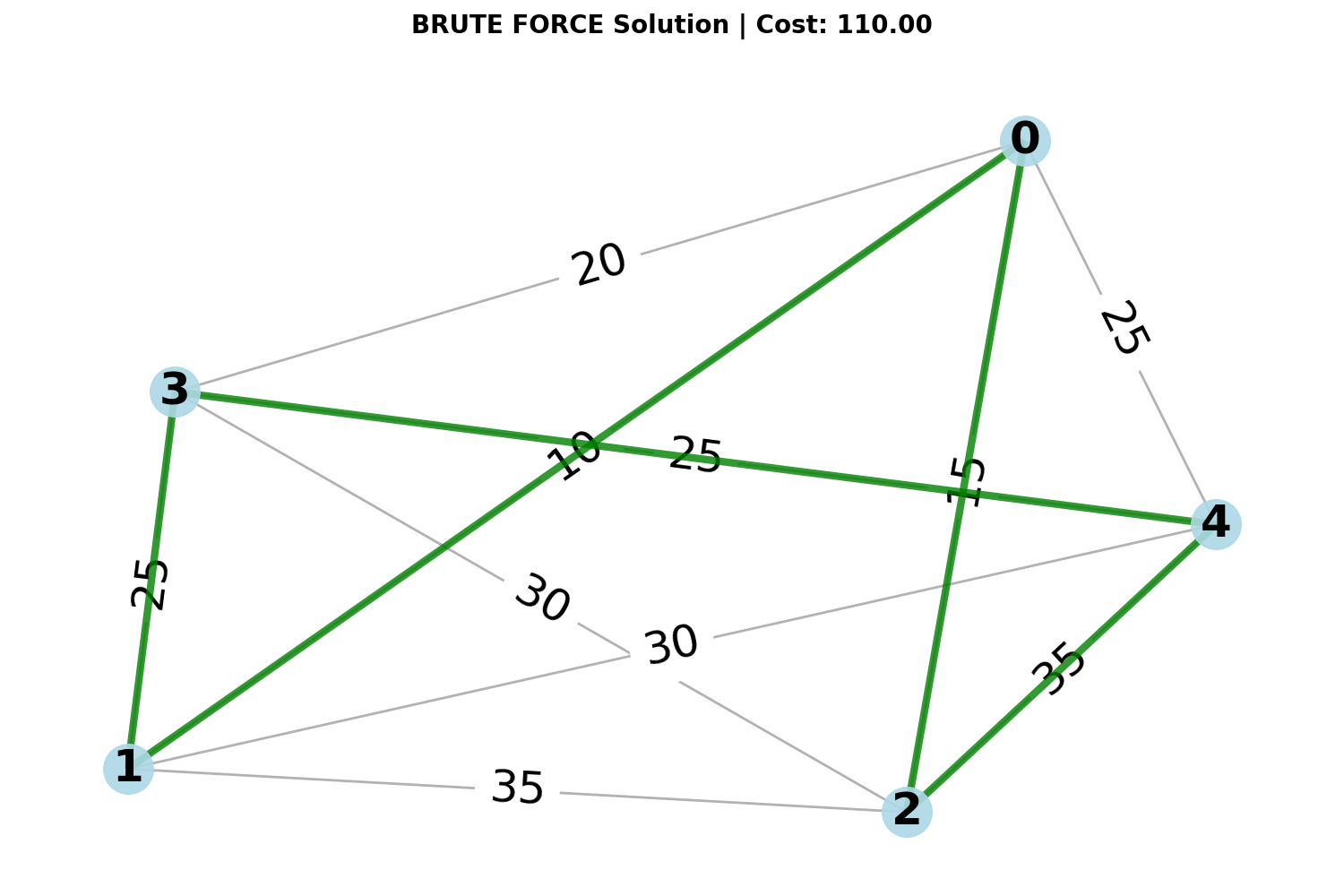}
        \caption{The Brute-Force algorithm, by enumerating all possible tours, finds the true optimal solution for this graph, establishing a benchmark cost of 110. The resulting path, shown below, serves as the ground truth for evaluating the other methods.}
        \label{fig:image1}
    \end{subfigure}
    \hfill 
    \begin{subfigure}{0.48\textwidth}
        \centering
        \includegraphics[width=\linewidth]{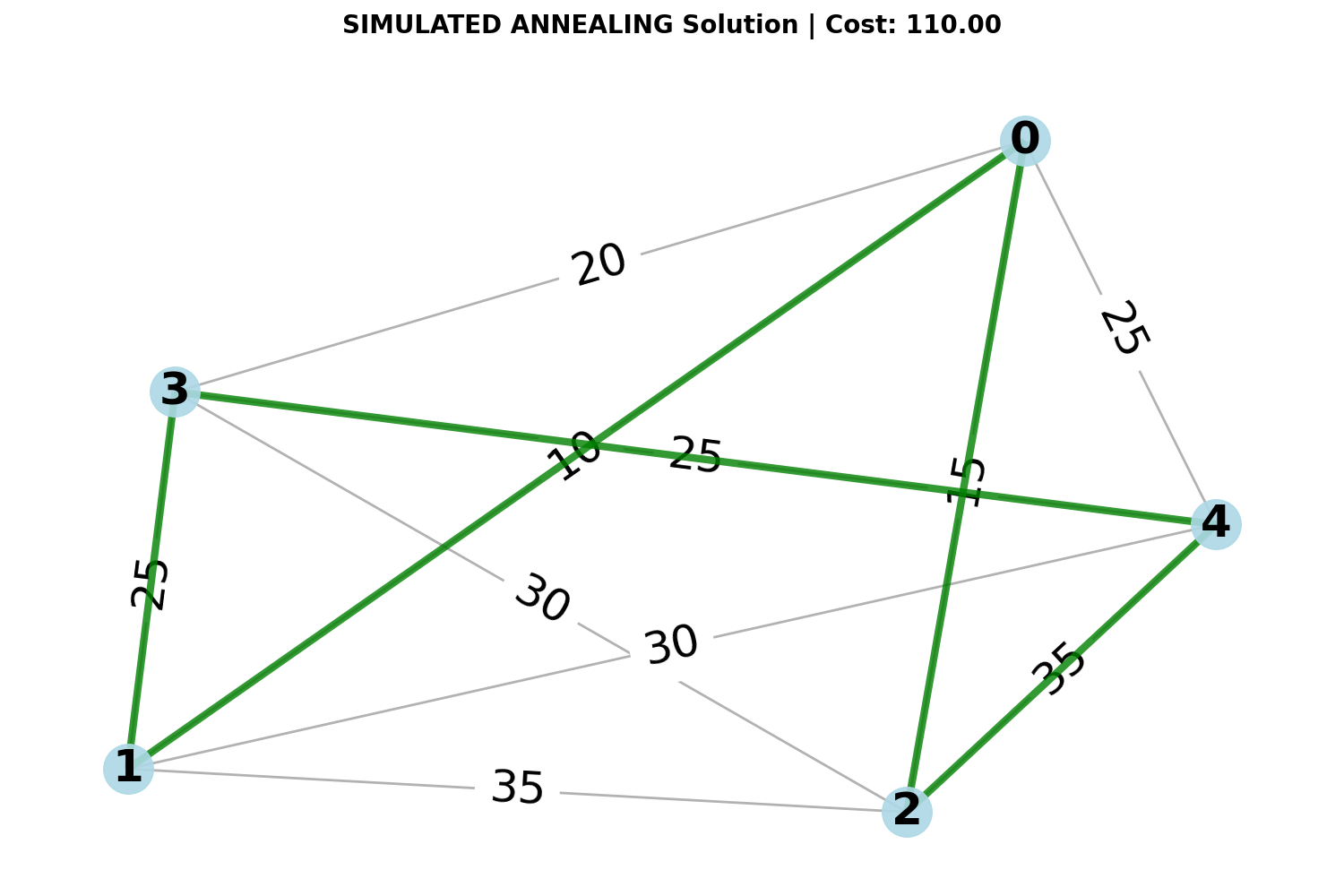}
    \caption{The Simulated Annealing algorithm also identifies this exact optimal tour, with $5001$ iterations. Despite being heuristic, the algorithm found the best path, underscoring its strength in delivering high-quality solutions for a fraction of the computation.}
        \label{fig:image2}
    \end{subfigure}

    \vspace{1em} 

    \begin{subfigure}{0.48\textwidth}
        \centering
        \includegraphics[width=\linewidth]{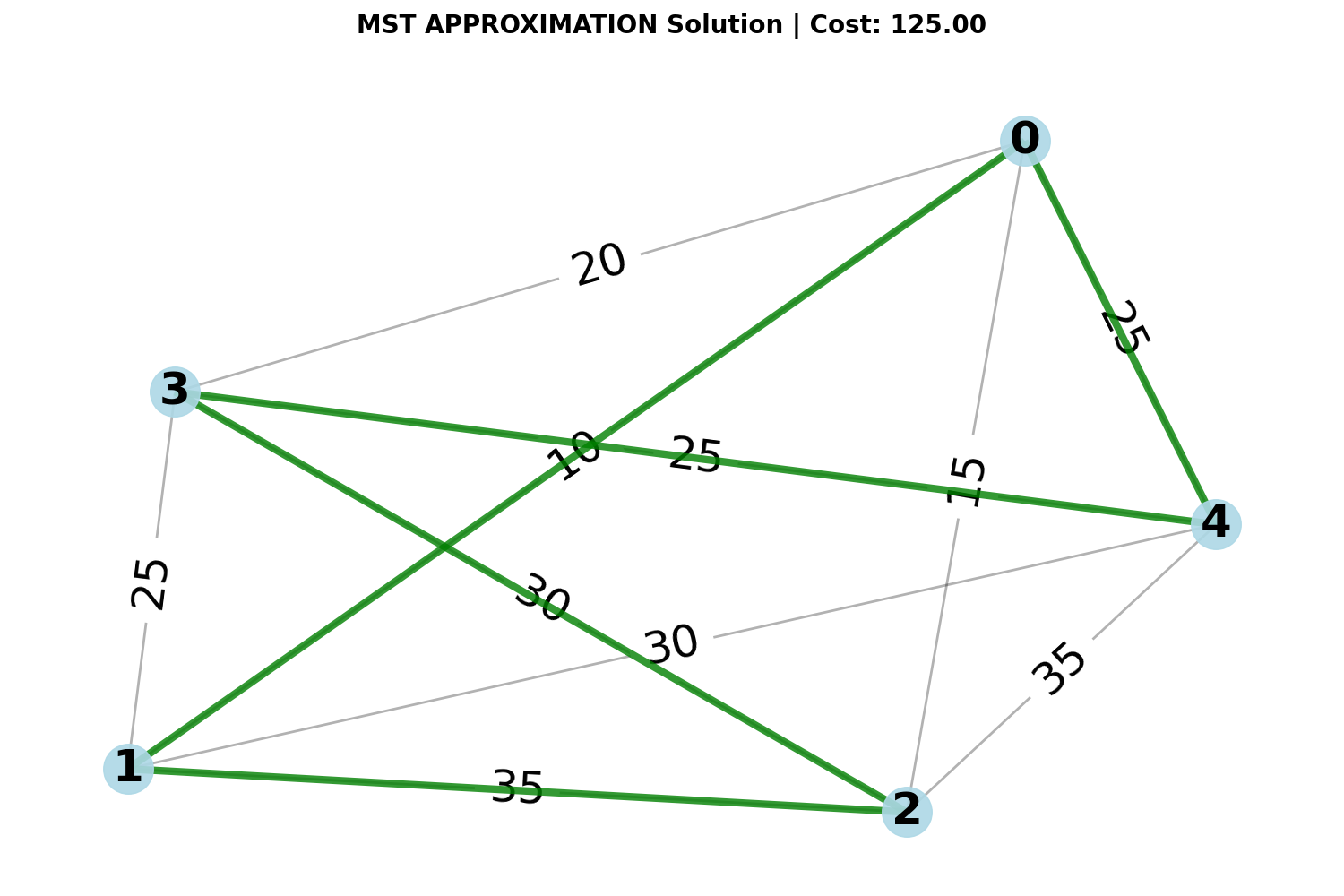}
    \caption{The MST 2-approximation algorithm yields a valid but longer tour with a total cost of 125. Although, the algorithm provides a fast, scalable solution with a guaranteed performance bound, this example illustrates how the resulting path can deviate from the true optimum.}
        \label{fig:image3}
    \end{subfigure}
    \hfill 
    \begin{subfigure}{0.48\textwidth}
        \centering
        \includegraphics[width=\linewidth]{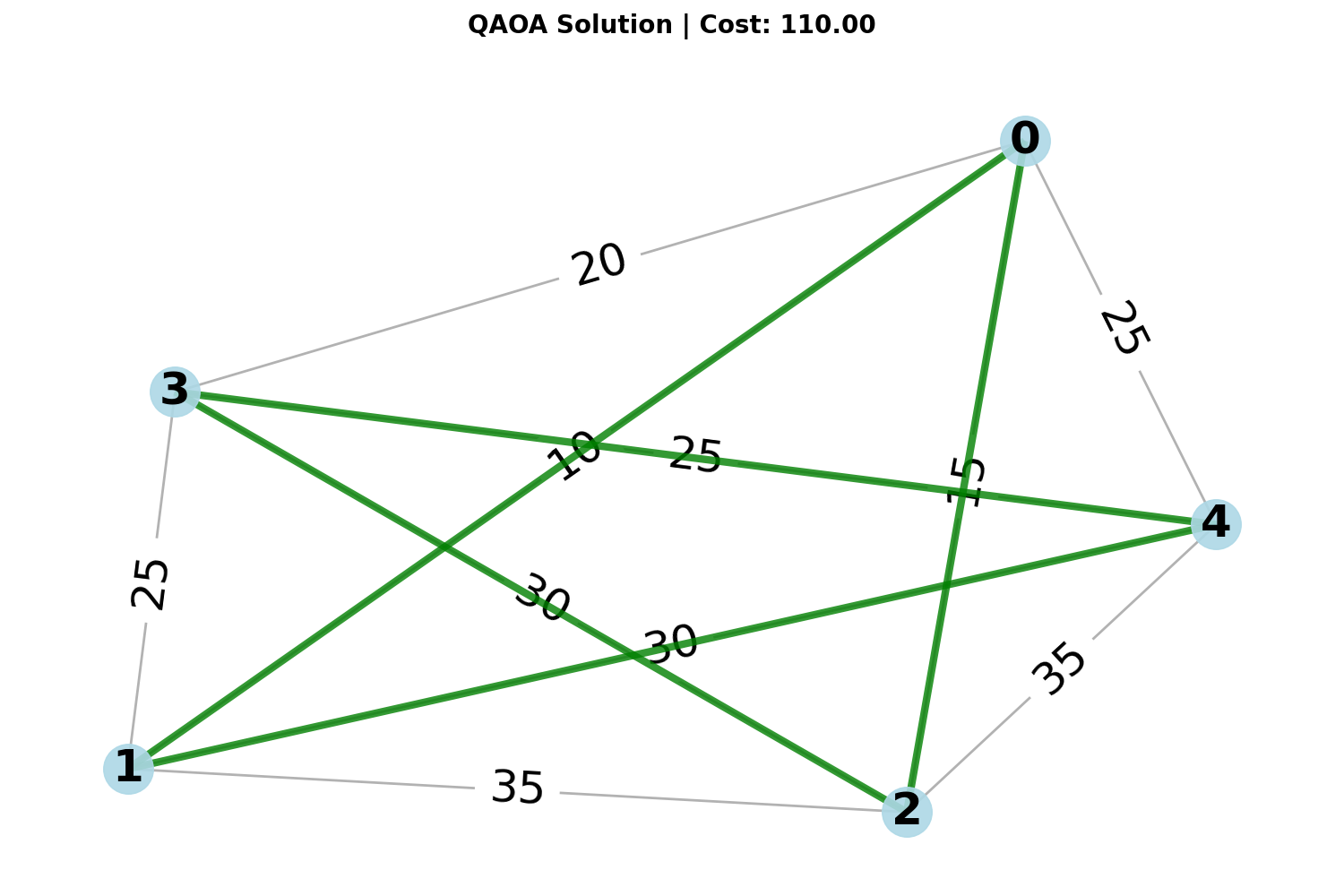}
    \caption{The QAOA (exact) implementation found an optimal tour with a cost of 110, but with an increased overhead in terms of memory and execution time. This result visually reinforces the challenges associated with the QUBO-based formulation for the TSP; even on a small 5-city problem.}
        \label{fig:image4}
    \end{subfigure}

    \caption{Paths found by each algorithm on a 5-city, fully connected graph.}
    \label{fig:2x2grid}
\end{figure*}

In summary, the results show a clear trade-off between optimality, performance, and solution quality. While brute force guarantees optimality, it is not scalable. Simulated annealing emerges as a highly effective method, achieving an excellent balance by consistently finding near-optimal solutions with minimal computational cost. Despite being the fastest algorithm tested, the MST approach is inherently bounded by its theoretical 2-approximation limit, which constrains its solution quality for the TSP. In contrast, the QAOA implementation is not always optimal, and the increased overhead in terms of space and time does not help its case. This is consistent with reports that highlight the difficulty in tuning QUBO-based approaches to reliably find high-quality solutions \cite{smithmiles}.

\section{Conclusions \& Discussion}
\label{sec:concl}

Our comparative study provides concrete, implementation-level evidence that supports the critical perspective on QUBO-based quantum optimizers for the TSP, as articulated by recent critical reviews in the field. The results show that a simulated QAOA, hampered by a classical computation bottleneck and the complexities of its underlying formulation, as of today may not be competitive with a standard classical heuristic like Simulated Annealing in either scalability or solution quality. These findings lead us to concur with the conclusion of \cite{smithmiles}: the path to demonstrating quantum advantage for the TSP is unlikely to be through general-purpose QUBO solvers. The challenges of an inefficient formulation, sensitive parameter tuning, and the formidable performance of highly optimized classical algorithms appear immense. Even though some novel and creative approaches like the single qubit formulations \cite{goswami2024solvingtravellingsalesmanproblem} exist, they face their own challenges with scalability, as well as some concerns and comments have been raised from a counter-proposal based on quantum genetic optimization \cite{10.1145/3712255.3734276}.

Across execution time, solution quality, and memory, the benchmarks expose a consistent separation between the classical and quantum pipelines. Trotterized QAOA extended simulation to seven cities but at the price of larger quality scatter and no clear monotonic improvement from increasing \texttt{num\_approx} at fixed depth $p=2$, indicating that discretization error and shallow circuits limit reliability before problem size does. The exact-versus-Trotterized backend comparison further showed that engineering choices matter as much as algorithm choice: exact evolution escalates in both time and memory because the full unitary must be decomposed, whereas Trotterization trades that bottleneck for deeper circuits and higher \texttt{num\_approx}-dependent resource use without restoring classical-level efficiency.

While our results cast doubt on the viability of this specific QAOA implementation, they also help to sharpen the open questions and future directions for the field. Rather than signaling an end to the quest for quantum solutions for the TSP, our work helps to clarify which paths may be more fruitful. This raises several important questions for future research.

\subsection{Open Questions and Future Directions}

\begin{itemize}
    \item \textbf{How can we move beyond QUBO?} A central argument is that the QUBO  formulation itself is a major roadblock for the TSP, creating an unnecessarily complex and difficult optimization landscape. The challenges faced by QUBO-based quantum methods today echo the historical struggles of similar approaches with Hopfield neural networks decades ago \cite{smithmiles}. This leads to a critical open question: \textit{What non-QUBO quantum algorithms or alternative problem mappings could be developed for the TSP?} Future work should explore novel formulations that might more naturally encode the TSP’s constraints and combinatorial structure, freeing them from the limitations that have hampered progress.

    \item \textbf{What is the true potential of hybrid quantum-classical models?} Perhaps the most promising avenue for near-term advantage lies not in purely quantum solvers, but in targeted hybrid approaches. Instead of attempting to replace classical solvers entirely, a key open question is: \textit{Can quantum subroutines be used to accelerate the most computationally difficult components of the best classical algorithms?} This is supported by recent advancements in large-scale QUBO solvers like VeloxQ \cite{veloxq}, which utilizes a physics-inspired methodology to achieve remarkable scalability that allows for solving problems with up to $2\times10^{8}$ variables, and is designed to be readily integrated into hybrid quantum-classical workflows to maximize the strengths of both technologies. Even though techniques like VeloxQ exist, demonstrating a practical speed-up for these general sub-problems remains a significant challenge.

    \item \textbf{How do we establish fair and meaningful benchmarks?} The field requires a more rigorous and standardized approach to benchmarking. A crucial direction for the community is to answer: \textit{What constitutes a fair and comprehensive benchmark for TSP solvers?} This involves developing evaluation methodologies that compare performance across different hardware and formulations, considering runtime, solution quality, and scalability on a diverse and challenging set of problem instances that reflect real-world difficulty.

    \item \textbf{What about pure quantum approaches?} One could argue to focus exclusively on purely quantum implementations to circumvent the limitations identified in hybrid and QUBO-based methods. However, current quantum hardware constraints such as, limited qubit counts, high error rates, and restricted connectivity, create fundamental barriers to fair evaluation. As quantum hardware matures, future research should revisit pure quantum approaches with more capable devices, potentially revealing advantages that are currently obscured by technological limitations.

\end{itemize}

By providing a grounded case study, we hope our work encourages the research community to shift focus from general-purpose QUBO solvers towards these more targeted and promising research questions, paving a more productive path toward understanding the potential for quantum advantage in combinatorial optimization.

Our open-source repository also includes a complete quantum implementation configured for execution on IBM Quantum hardware.
Empirical evaluation on physical backends lies outside the scope of this simulation-based study and will be addressed in future work.\\

\noindent \textbf{Use of Generative AI: } LLMs were used to structure sentences for clarity and to suggest alternative phrasings for the manuscript. The final manuscript was carefully reviewed and corrected by the authors to ensure precision.

%% file: ack.tex
\section*{Acknowledgment}

This work was supported by the Natural Sciences and Engineering Research Council of Canada (NSERC)
through an Undergraduate Student Research Award (USRA) to Krit Grover.

Numerical experiments and more advanced computations were performed on the Trillium supercomputers at the SciNet HPC Consortium. SciNet is funded by Innovation, Science and Economic Development Canada; the Digital Research Alliance of Canada; the Ontario Research Fund: Research Excellence; and the University of Toronto.